\documentclass{aa}
\usepackage[varg]{txfonts}
\usepackage{graphicx,epsfig,fancyhdr,rotating,amsmath,natbib}

\def\etal{{et al. }}

\def\3o{O~{\sc ii}}
\def\4o{O~{\sc iv}}

\usepackage[none]{hyphenat}

\begin{document}

\title{Diagnosing transient ionization in dynamic events}

\author{J.G. Doyle$^{1}$, A. Giunta$^{2,3}$, M.S. Madjarska$^{1}$, H. Summers$^{2,3}$,  
M. O'Mullane$^{2,3}$, \and A. Singh$^{1,4}$}
\offprints{madj@arm.ac.uk}
\institute{
$^{1}$Armagh Observatory, College Hill, Armagh BT61 9DG, N. Ireland\\
$^2$Department of Physics, University of Strathclyde, 107 Rottenrow, Glasgow, G4 0NG, Scotland\\
$^3$Space Science and Technology Department, STFC Rutherford Appleton Laboratory, Chilton, 
Didcot, Oxfordshire, OX11 0QX, UK.\\
$^4$Dept.of Physics and Electronics, Deen Dayal Upadhyaya College, University of Delhi, India}

 \date{Received date, accepted date}

\abstract
{}
{The present study aims to provide a diagnostic line ratio that will enable the observer to determine 
whether a plasma is in a state of transient ionization. }
{We use the Atomic Data and Analysis Structure (ADAS) to calculate line contribution functions for two 
lines, Si~{\sc iv}~1394~\AA\ and O~{\sc iv}~1401~\AA, formed in the solar transition region. The generalized 
collisional-radiative theory is used. It includes all radiative and electron collisional processes, except for photon-induced processes. State-resolved direct
ionization and recombination to and from the next ionization stage are also taken into account.}
{For dynamic bursts with a decay time of a few seconds, the Si~{\sc iv}~1394~\AA\ line can be 
enhanced by a factor of 2--4 in the first fraction of a second with the peak in the line contribution 
function occurring initially at a higher electron temperature due to transient ionization compared to ionization equilibrium
conditions. On the other hand, the O~{\sc iv}~1401~\AA\ does not show such any enhancement. Thus the ratio of
these two lines, which can be observed with the Interface Region Imaging Spectrograph, 
can be used as a diagnostic of transient ionization.}
{We show that simultaneous high-cadence observations of two lines formed in the solar transition region may be used as 
a direct diagnostic of whether the observed plasma is in transient ionization. The ratio of these two lines can
change by a factor of four in a few seconds owing to transient ionization alone.}

\authorrunning{Doyle, M. S. et al.}
\titlerunning{Diagnosing Transient Ionization in Dynamic Events}

\maketitle

\section{Introduction}

All spectral lines have encoded information that allows the observer 
to diagnose important physical parameters of the underlying plasma. For example, 
forbidden or inter-system lines allow an evaluation of the plasma's electron density, while resonance lines
allow the observer to diagnose the electron temperature. With sufficient spectral resolution, all spectral 
lines can be used to give the observer information on plasma flows and/or turbulent motions. To diagnose
a plasma's state of ionization however, requires both high-cadence data and suitable spectral lines.

Various authors have presented simulated spectra based on transient ionization \citep{1980A&A....87..261M, 2008ApJ...684..715R}
 for comparison with observational data. In a recent paper, Doyle \etal (2012) 
discussed the diagnostic potential of high-cadence ultraviolet spectral data when transient ionization 
is considered. The above paper used high-cadence spectral line data from the Solar Maximum 
Mission (SMM) observed in O~{\sc v}~1371~\AA\ which allowed the authors to measure electron densities and temperatures 
during the early stages of a feature's evolution, something that is not currently 
possible. The high cadence UV spectrometer on SMM allowed observations of selected spectral lines with 
a subsecond cadence.

The forthcoming observations from the  Interface Region Imaging Spectrograph\footnote{http://iris.lmsal.com/} will once again enable 
high quality observations of lines formed in the solar transition region.
What we would like to have is a simple line ratio consisting of one line that shows a response to transient 
ionization and another line that is not responsive to transient ionization. Then, provided we can monitor 
these lines with sufficient cadence, any fast ($\approx$ 0.5~s) increase in the ratio must be due to non-equilibrium
conditions.  Here, we look at the response of two such lines, Si~{\sc iv}~1394~\AA\ and 
O~{\sc iv}~1401~\AA. The Si~{\sc iv}~1394~\AA\ line is in the ground spin system (3s $^2$S$_{1/2}$ -- 3p $^2$P$_{3/2}$).  
The character of the excitation cross-sections (dipole and non-spin change) and access to 
higher n-shell/cascade will result in a line enhancement. However, the O~{\sc iv}~1401 line, by contrast, 
has an upper state in the quartet spin system (2s$^2$2p $^2$P$_{3/2}$ -- 2s2p$^2$ $^4$P$_{5/2}$) which means that 
access to this spin system is driven by cross-sections that decrease with temperature, hence little or no line 
enhancement, therefore, transient ionization under-fills the population structure of the alternate spin systems 
from the ground. By contrast, the ground spin system population structure gets enhanced (due to the exponential 
factor in the rate coefficient) with increasing electron temperature.

\begin{figure}[ht!]
 \centering
 \vspace{14.5cm}
\includegraphics{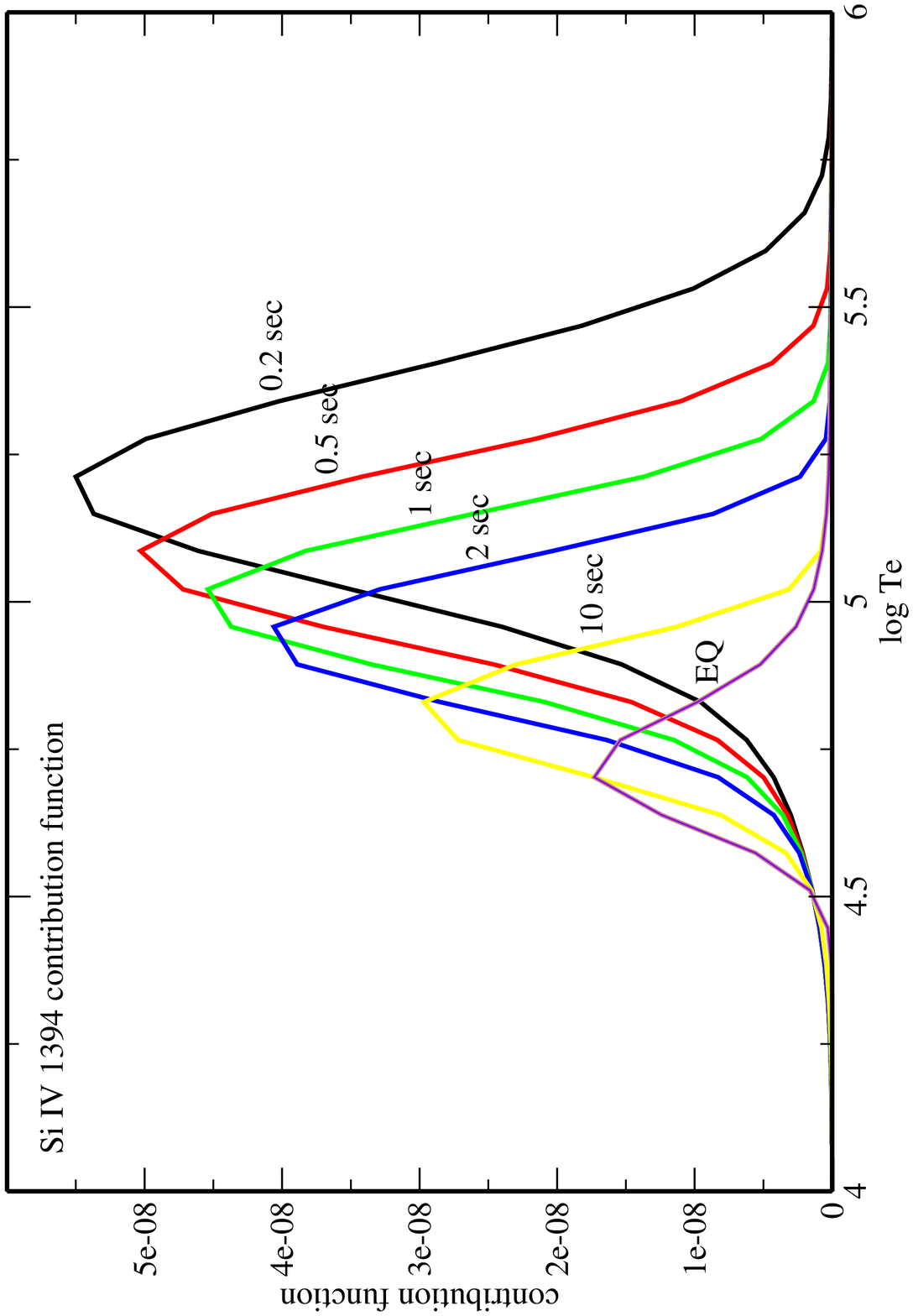}
\includegraphics{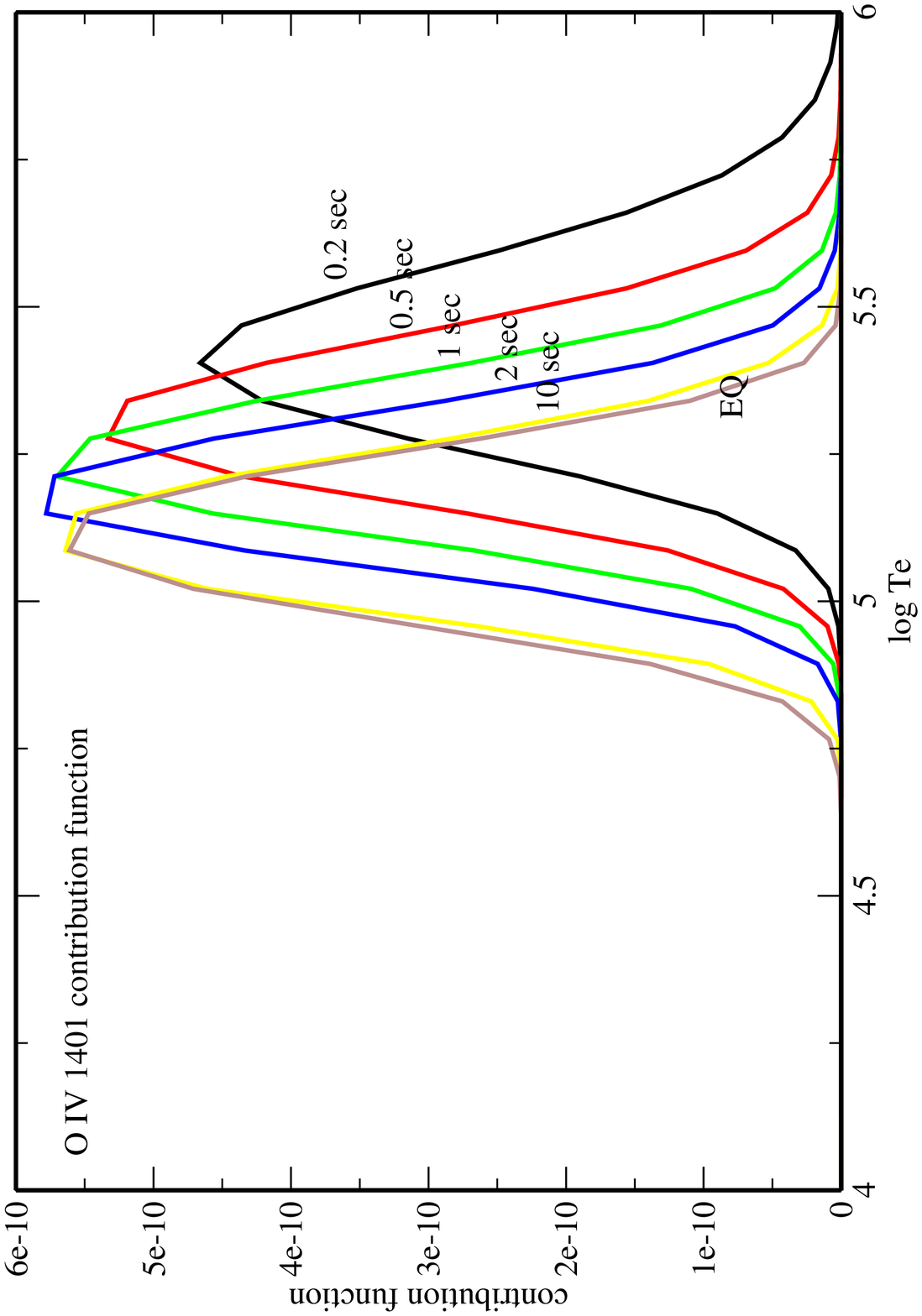}
\vspace*{-1cm}
 \caption{Line contribution function for Si~{\sc iv}~1394~\AA\ and O~{\sc iv}~1401~\AA\ in cm$^{-3}$ s$^{-1}$ at ionization
 equilibrium and for transient ionization. The electron density was fixed at $10^{10}$~cm$^{-3}$, with 
 an initial starting temperature of 30\,000K.}
 \label{fig1}
 \end{figure}
\section{Transient Ionization}

The atomic structure of atoms and ions is in principle an infinite assembly of levels with an infinite
number of reactions between them, however, simplifying assumptions about the nature of the plasma, its dynamic
character and the relative importance of the various reactions have to be made. As shown in the recent paper 
by Doyle \etal (2012), the generalized collisional-radiative theory (GCR) is the most appropriate approach. 
The above theory, plus ionization, recombination, excitation and radiative decay data are all as implemented 
in ADAS, the Atomic Data and Analysis Structure\footnote{http://www.adas.ac.uk/manual.php}. 
A good description of this is given in the Appendix of \citet{2001ApJ...547.1116L}. Below we first discuss 
the various time constants and their relative importance.

\subsection{Time constants}

The lifetimes of the various states of ions and 
electrons in a plasma too radiative or collisional processes divide into two groups. The first is
the intrinsic group, comprising purely of atomic parameters, and includes the metastable radiative
decay time, $\tau_m$, the ordinary excited-state radiative decay time $\tau_o$, and auto-ionizing-state decay time, 
$\tau_a$. These are generally ordered as $\tau_a << \tau_o << \tau_m$.

The second is the extrinsic group, which depends on plasma conditions, especially density. It 
includes free particle thermalization including electron-electron $\tau_{e-e}$, ion-ion $\tau_{i-i}$, 
and ion-electron $\tau_{i-e}$, charge-state change (ionization $\tau_{ion}$ and recombination 
$\tau_{rec}$) and redistribution of population amongst excited ion states ($\tau_{red}$). Their 
order is
$\tau_{ion,rec} >> \tau_{i-e} >> \tau_{i-i} >> \tau_{e-e}$.
More details on the above may be found in Summers \etal (2006).

From a dynamic point of view, the intrinsic and extrinsic groups are to be compared with
each other and with timescales, $\tau_{plasma}$, representing plasma ion diffusion across temperature
or density scale lengths, the relaxation times of transient phenomena, and observation times. For
astrophysics plasmas,
$\tau_{plasma} \approx \tau_g \approx \tau_m >> \tau_o >> \tau_{e-e}$,
where $\tau_g$ represents the relaxation time of the ground-state populations of ions. 

These time scales imply that the dominant population of impurities in the plasma are those of the ground and
metastable states of the various ions. In a transient event, the dominant populations evolve on 
time scales similar to the plasma diffusion time scales and so should be modelled dynamically.
From a theoretical point-of-view, non-equilibrium occurs when $\tau_{plasma} \le
\tau_{ion,rec}=1/[N_e(S^{(z\rightarrow z+1)}+\alpha^{(z+1\rightarrow z)})]$ \citep{1965pdt..conf..201M}, 
where $N_e$ is the electron density, and $S^{(z\rightarrow z+1)}$ and $\alpha^{(z+1\rightarrow z)}$ 
are the effective ionization and recombination coefficients (see Section 2.2). This means that in 
a transient event, the plasma time scale ($\tau_{plasma}$) is shorter than the atomic ionization/recombination 
time scale ($\tau_{ion,rec}$).

\subsection{GCR} 

Here, we calculate the contribution functions for Si~{\sc iv}~1394~\AA\ and O~{\sc iv}~1401~\AA\ 
using the GCR theory. Each ion 
in an optically thin plasma is described by a complete set of levels with collisional and 
radiative couplings between them. All radiative and electron collisional processes are 
included, except for photon-induced processes. In addition, state-resolved direct
ionization and recombination to and from the next ionization stage are also taken
into account.  

The emissivity of a spectral line is given by
 \begin{equation}
\varepsilon_{j\rightarrow i}= A_{\rm el} {{N_{\rm H}}\over {N_{\rm e}}} N^2_{\rm e} G^{(\it z)}_{j\rightarrow i} (T_{\rm e},N_{\rm e},t)
\label{line_emissivity_2}
\end{equation}
where $A_{\rm el}=N^{(\it Z)}/N_{\rm H}$ is the abundance of the element $\it Z$ relative to hydrogen, $N_{\rm H}/N_{\rm e}$, tabulated by \citet{1975A&A....40...63M}, and $G^{(\it z)}_{j\rightarrow i} (T_{\rm e},N_{\rm e},t)$ is the 
time-dependent contribution function defined as
 \begin{equation}
G^{(\it z)}_{j\rightarrow i} (T_{\rm e},N_{\rm e},t) = \mathcal{PEC}^{\rm (exc, \it z)}_{j\rightarrow i} {{N^{\it (z)}(t)}\over {N^{\it (Z)}}} + \mathcal{PEC}^{\rm (rec,\it z)}_{j\rightarrow i} {{N^{\it (z+1)}(t)}\over {N^{\it (Z)}}}
\label{time_contr_func}
\end{equation}

\noindent
where ${PEC}^{\rm (exc, \it z)}_{j\rightarrow i}=A_{j\rightarrow i} \mathcal{F}^{\rm (exc, \it z)}$ is the excitation 
photon emissivity coefficient, and ${PEC}^{\rm (rec, \it z)}_{j\rightarrow i}=A_{j\rightarrow i} \mathcal{F}^{\rm (rec, \it z)}$
the recombination photon emissivity coefficient. 

In the above, $A_{j\rightarrow i}$ is the radiative transition probability, $N^{\it (z)}$ and $N^{\it (z+1)}$ are 
the population densities of the ground states of the ions of charge 
$\it z$ and $\it z+1$ and $\mathcal{F}^{\rm (exc, \it z)}$, and
$\mathcal{F}^{\rm (rec,\it z+1)}$ are the effective contributions to the
population of the upper excited state $i$. 

In a time-dependent plasma model, the line emissivity is no longer a unique
function of the local temperature and density conditions, but it depends on the
past history of the temperature, density, and state of ionization of the
plasma. Therefore, the assumption of ionization equilibrium in calculating the 
ionization balance is not appropriate and time-dependent fractional abundances must be determined.
The time dependence of ionization stage populations, $N^{(\rm z)}$, leads to the following equation:
\begin{equation}
{{dN^{(\it z)}} \over {dt}} = N_{\rm e} [S^{(\it z-1)}N^{\it (z-1)} + (S^{(\it z)}+\alpha^{\it (z)})N^{(\it z)} + \alpha^{(\it z+1)}N^{(\it z+1)}]
\label{ion_time_evol}
\end{equation}
where $S$ and $\alpha$ are the collisional-dielectronic ionization and recombination coefficients. They 
give the contribution to the growth rates for the ground state population due to the effective 
ionization, which includes direct and excitation/auto-ionization contributions, and the effective 
recombination, which includes radiative, dielectronic, and three-body contributions. The values 
of these coefficients, currently within the ADAS database, have been obtained following the GCR 
approach as described in Summers \etal (2006). The solution of Equation (\ref{ion_time_evol}) is such 
that the number density of the element of a nuclear charge $\it Z$, $N^{(\it Z)}$, is equal to 
$\sum^{\it Z}_{\it {z=0}}N^{(\it z)}$. The time-dependent fractional abundances $N^{(\it z)}(t)/N^{(\it Z)}$ are 
calculated as follows: the code derives the solution for a range of fixed plasma 
electron temperature and density pairs, starting from an initial population distribution 
$N^{(\it z)}(t=0)/N^{(\it Z)}$ using an eigenvalue approach. At temperatures close to ionization 
equilibrium, the contribution of recombination to the emissivity is small; however, this contribution
becomes large in non-equilibrium conditions. Further details on the above may be
found in Doyle \etal (2012) and references therein. Here we use an initial starting temperature of
30\,000K and a fixed electron density of 10$^{10}$ cm$^{-3}$. We then calculate the ionization fraction and 
the resulting line contribution function for different relaxation times, ranging from 0.2 sec after the initial
start (i.e., in a highly transient state) to 10 sec (close to ionization balance) and finally when the 
plasma is in ionization equilibrium.

\section{Results \& discussion}

In Fig.~1, we give the line contribution function for different relaxation times, that are shorter 
than the time for the plasma to reach ionization equilibrium. This figure shows that the peak of the line 
contribution function for both lines occurs at a higher temperature for short relaxation times. However, for 
O~{\sc iv}~1401~\AA, transient ionization does not produce a flux increase over the one for ionization 
equilibrium, while the Si~{\sc iv}~1394~\AA\ line increases over a factor of three.  
The reason these two lines behave in a different manner under transient ionization is due to
their formation process as outlined in Section 1. For O~{\sc iv}~1401~\AA\, at an electron density 
of $10^{10}$~cm$^{-3}$, it takes $\approx$ 10 sec for the line to reach ionization equilibrium, while 
Si~{\sc iv}~1394~\AA\ takes $\approx$ 100 sec to reach equilibrium.

Figure 1 shows that the ratio of Si~{\sc iv}~1394 to O~{\sc iv}~1401 ranges from $\approx$ 30 in ionization 
equilibrium to $\approx$ 120 for highly transient conditions (0.2 s). Multiplying by the relative elemental 
Si/O coronal abundance \citet{2008uxss.book.....P}  gives a ratio ranging from 3.3 to 12.9. The \citet{2001A&A...375..591C}
SUMER spectral atlas gives the observed ratio for these lines as 4.6 in a sunspot and  6.9 in the quiet Sun. 
The above observational ratios are in general agreement (to within a factor of two) with the present 
Si~{\sc iv}~1394/O~{\sc iv}~1401 ratio when assuming ionization equilibrium. However, with a different DEM distribution, 
different abundances will change the ratio but the important point to note by the observer is not 
the absolute ratio but whether the observed ratio would change quickly. 
As a result, if observing these two lines simultaneously with, say, a 0.5~s cadence, then any 
sudden increase by a factor of 2--3 in the line ratio must be due to transient ionization. Such observations 
will therefore enable an evaluation of the plasma's electron density as outlined \citet{2012SoPh..280..111D}.
\begin{acknowledgements} Research at the Armagh Observatory is grant-aided by
the N. Ireland Dept. of Culture, Arts and Leisure. We thank UK STFC for support 
via ST/J001082/1 and the Leverhulme Trust.
\end{acknowledgements}

\bibliographystyle{aa}

\begin{thebibliography}{8}
\expandafter\ifx\csname natexlab\endcsname\relax\def\natexlab#1{#1}\fi

\bibitem[{{Curdt} {et~al.}(2001){Curdt}, {Brekke}, {Feldman}, {Wilhelm},
  {Dwivedi}, {Sch{\"u}hle}, \& {Lemaire}}]{2001A&A...375..591C}
{Curdt}, W., {Brekke}, P., {Feldman}, U., {et~al.} 2001, \aap, 375, 591

\bibitem[{{Doyle} {et~al.}(2012){Doyle}, {Giunta}, {Singh}, {Madjarska},
  {Summers}, {Kellett}, \& {O'Mullane}}]{2012SoPh..280..111D}
{Doyle}, J.~G., {Giunta}, A., {Singh}, A., {et~al.} 2012, \solphys, 280, 111

\bibitem[{{Lanza} {et~al.}(2001){Lanza}, {Spadaro}, {Lanzafame}, {Antiochos},
  {MacNeice}, {Spicer}, \& {O'Mullane}}]{2001ApJ...547.1116L}
{Lanza}, A.~F., {Spadaro}, D., {Lanzafame}, A.~C., {et~al.} 2001, \apj, 547,
  1116

\bibitem[{{McWhirter}(1965)}]{1965pdt..conf..201M}
{McWhirter}, R.~W.~P. 1965, in Plasma Diagnostic Techniques, ed. R.~H.
  {Huddlestone} \& S.~L. {Leonard}, 201

\bibitem[{{McWhirter} {et~al.}(1975){McWhirter}, {Thonemann}, \&
  {Wilson}}]{1975A&A....40...63M}
{McWhirter}, R.~W.~P., {Thonemann}, P.~C., \& {Wilson}, R. 1975, \aap, 40, 63

\bibitem[{{Mewe} \& {Schrijver}(1980)}]{1980A&A....87..261M}
{Mewe}, R. \& {Schrijver}, J. 1980, \aap, 87, 261

\bibitem[{{Phillips} {et~al.}(2008){Phillips}, {Feldman}, \&
  {Landi}}]{2008uxss.book.....P}
{Phillips}, K.~J.~H., {Feldman}, U., \& {Landi}, E. 2008, {Ultraviolet and
  X-ray Spectroscopy of the Solar Atmosphere} (Cambridge University Press)

\bibitem[{{Reale} \& {Orlando}(2008)}]{2008ApJ...684..715R}
{Reale}, F. \& {Orlando}, S. 2008, \apj, 684, 715

\end{thebibliography}

\end{document}